# Transport and Capacitance properties of Charge Density Wave in few layer 2H-TaS$_2$ Devices

CAO Yu-Fei(曹玉飞)[1], CAI Kai-Ming(蔡凯明)[1], LI Li-Jun(黎丽君)[2], LU Wen-Jian(鲁文建)[2]
SUN Yu-Ping(孙玉平)[2], WANG Kai-You(王开友)[1*]

[1]*The State Key Laboratory of Superlattices and Microstructure, Institute of Semiconductors, Chinese Academy of Sciences, Beijing 100083, China*

[2] *Key Laboratory of Materials Physics, Institute of Solid State Physics, Chinese Academy of Sciences, Hefei 230031, China*

## Abstract

We carefully investigated the transport and capacitance properties of few layer charge density wave (CDW) 2H-TaS$_2$ devices. The CDW transition temperature and the threshold voltage vary from device to device, which is attributed to the interlayer interaction and inhomogeneous local defects of these micro-devices based on few layer 2H-TaS$_2$ flakes. Semiconductivity rather than metallic property of 2H-TaS$_2$ devices was observed in our experiment at low temperature. The temperature dependence of the relative threshold voltage can be scaled to $(1-T/T_r)^{0.5+\delta}$ with $\delta = 0.08$ for the different measured devices with presence of the CDWs. The conductance-voltage and capacity-voltage measurements were performed simultaneously. At very low ac active voltage, we found that the hysteresis loops of these two measurements exactly match each other. Our results point out that the capacity-voltage measurements can also be used to define the threshold depinning voltage of the CDW, which give us a new method to investigate the CDWs.

[*] Corresponding author's E-mail: kywang@semi.ac.cn

## Introduction

Layered transition-metal dichalcogenides chalcogenides (TMDC) exhibit rich physical properties. Charge density wave (CDW) and superconductivity coexist in most these quasi low dimensional materials such as 2H–TaSe$_2$, 2H–NbSe$_2$, 2H–TaS$_2$, 4Hb–TaS$_2$, and 4Hb–TaSe$_2$.[1-3] The charge-density wave (CDW) order originates from the instability of low-dimensional electronic systems due to the nonzero electron-phonon coupling. The hexagonally ordered sheets of tantalum atoms in 2H-TaS$_2$ are sandwiched between sheets of sulfur atoms, which undergoes a CDW transition at 75 K and a superconducting transition at 0.8 K .[4,5] The CDW in 2$H$-TaS$_2$ has been expected predominantly from the hybridization of Ta atoms.[6] Once the measured temperature falls in the temperature range of existing CDW phase. At very small electrical field, only the single particle current flows in the device and the CDW remains pinned to the defects. However, above a certain threshold electrical field, the CDW was depinned and provided a collective charge current adding to the single particle current associating with a sudden current intensity rising.[7,8]

Although the CDW in the 2H-TaS$_2$ have been extensively investigated, there is only very a few reports based on few layers of these system.[9-12] In their reports, the disappearance of the charge density wave phase anomaly at low temperature was observed.[11,12] The transport and optical methods were widely used to investigate the CDW.[10,13,14] However, to our knowledge, there is not any research on capacitance measurements about the CDW yet. In this paper, we not only investigated the temperature dependence of the current-voltage characteristics and also the temperature dependence of the current-capacitance characteristics. At very low ac active voltage, we found that the hysteresis loops of the capacitance measurements exactly matches with that of the current voltage measurements. It suggests that the capacity-voltage measurements can also be used to define the threshold depinning voltage of the CDW.

## Experiments

The 2H–TaS$_2$ single crystal was grown using the NaCl/KCl flux method.[15] The structure and the transport properties of the bulk 2H–TaS$_2$ single crystal has been published elsewhere.[16] We pre-patterned the location marks using optical lithography on a SiO$_2$/Si substrate, then the ultrathin 2H-TaS$_2$ nanosheets was mechanically exfoliated on it,[11] where the thickness was characterized by atomic force microscopy (AFM). Using the location marks, the contacts were defined by e-beam lithography. Firstly 10 nm Ti and then 100 nm Au were deposited using thermal evaporator acting as source and drain electrodes. After lift-off and packaging, the transport and capacity properties of the fabricated devices at low temperatures were measured using Agilent B1500 in the Cryogenic free system.

**Results and Discussions:**

The thinnest 2H-TaS$_2$ nanosheet with thickness of ~ 2.11 nm, corresponding to 3 layers,[17] was measured by using AFM (Fig. 1 (a)). The schematic diagram and the optical image of one of our fabricated devices was shown in Fig.1 (b) and (c). The thickness of the devices shown in this paper is about 10 to 20 nm. When the thickness of TaS$_2$ below 5 nm, the fabricated devices show insulating behavior, which has also been observed by others.[11]

Fig. 2 (a) shows the current-voltage (I-V) loop characteristic for both device #A and #B. These two devices were made exactly in the same experimental procedures, where the only difference that the thickness of device #B is thinner than that of device #A.. The antisymmetrical hysteresis loops were observed between the positive and negative applied voltages for both devices. The current is nonlinearly rather than linearly increased with increasing the voltage before the hysteresis observed, which is not expected for the metallic property of bulk TaS$_2$. Also the linear I-V characteristic was observed by Ayari et al. even with only 8 nm thick of 2H-TaS$_2$.[11] In order to find the reason, we then carefully examined the I-V characteristics of device #A at different temperatures, which is shown in the inset of Fig.2 (a). The current linearly varies with changing the applied voltages at room temperature, which can rule out the possibility of the bad contacts. With decreasing the temperature till 180 K, the nonlinear I-V characteristic was shown up, which became more nonlinear with further decreasing the temperature. Thus it proved that the nonlinear I-V characteristic is originated from the thin TaS$_2$ side rather than the metal contacts. The possible reason of the nonlinear current voltage characteristic at low temperature could be: (1) The deformation of the structure during the exfoliation for the very thin device could result in the commensurate to incommensurate transition;[18,19] (2) the surface defect states from oxidation will result in Anderson localization and weaken the p-d hybrid interaction.[20] Future work need to be done to verify this origin.

We then carefully investigate the current-voltage loop characteristic at positive applied voltages for device #A and #B at different temperatures, which are shown in Figs. 2 (b) and (c). The current-voltage hysteresis loops were observed for both devices at low temperatures, which become narrower with increasing the temperatures. However, in some of our measured devices, there is not any hysteresis observed as shown in Figs. 2(b) and(c), which is similar to that of the previous reports in the few layer of 2H-TaS$_2$ with the disappearance of the CDW.[11,12] The observed hysteresis loops are attributed to the presence of the CDW phase, where the hysteresis loop is much sharper for the device #B. Furthermore, the CDW phase transition temperature for the device #A is 39 and 59 K for the device #B, while both of them are much lower than that of the bulk CDW phase transition temperature 75 K.[21] Thus we can conclude that the presence of the CDW phase may not only depends on the interlayer interaction, but also depends on the local electronic or defects states.

Comparing the details of Fig. 2 (b) and (c), the current firstly nonlinearly rather than linearly increases with increasing the electrical field at lower electrical field. This is very different to that of the room temperature and the bulk devices shown in the inset of Fig.2 (b), where the linearly Ohmic-law was obeyed. It indicates the barrier formed at the interface between the contacts and the 2H-TaS$_2$ flake, which results in a very low current intensity at small applied voltage, where the CDW remains pinned to the defects of the underlying lattice. When the applied voltage is larger than that of the threshold voltage, the CDW can depin from the defects and slides through the crystal associated with an extra collective charge current. It results in a sudden sharp increase of current intensity at the threshold voltage because of the collective charge transport contribution due to the moving CDW. The threshold voltage needed to depin the CDW at 2 K is about 0.54 V for the device #A and 0.88 V for the device #B. When the applied voltage sweeps back from the value above the threshold voltage, the CDW state can still persist with the voltage below the forward threshold voltage. The difference between the forward threshold voltage and backward recovering voltage is about 0.1 V for the device #B and 0.05 V for the device #A. Compared with the device #A, the electrical current of the device #B just before the threshold voltage is only 1% of that for the device #A. However, the electrical current of the device #B right above the threshold voltage is about 25% of that for the device #A. The only difference between these two device is that the thickness of device #B is about half of that device #A, which suggests that the interface barrier can more seriously affect the electrical transport in the thinner device. This also confirms that why the monolayer or double layer devices show insulating behavior.

By defining the depinning threshold voltage of the CDW at the peak of d$I$/d$V$ in forward bias, then the temperature dependence of the threshold voltage needed to depin the CDW can be obtained. In order to understand the inside physics of the electrical field driven CDW, the relative threshold voltage difference with decreasing temperature to depin the CDW can be written down as: $\Delta V = V_T - V_{Tr}$, where $V_T$ is the depinning threshold voltage at temperature $T$, and $V_{Tr}$ is the depinning threshold voltage at the CDW phase transition temperature. The temperature dependence of the $\Delta V$ for both devices are shown in Fig. 3 (a). Because the magnitude of $\Delta V$ is different for these two devices due to different pinning potential, it is hard to compare these two sets of data directly. In order to compare the temperature dependence of the relative $\Delta V$ for these two devices, we normalized $\Delta V$ to the voltage difference between 2 K and the CDW transition temperature. The normalized temperature dependence of $\Delta V/(V_{2K}-V_{Tr})$ for both devices are shown in the Fig. 3 (b). It is striking that the normalized two sets of data have the same trend. Thus we can conclude that both devices have the same scaling law for the temperature dependence of the relative depinning threshold voltage, indicating the same inside physics for both devices. Then the normalized curve is found to be proportional to $(1-T/Tr)^{0.5+\delta}$ with $\delta = 0.08$, where $T_r$ is the CDW phase transition temperature. An equation of the same form with $\delta = 0.0$ is also seen in the BCS theory for the superconductivity, which has also been widely used to describe the temperature dependence of the energy gap at the Brillouin boundary in CDW.[22-24]

Fig.4 shows the typical voltage-conductance and voltage-capacitance measured simultaneously with 5 MHz ac excitation $V_{ac}$ = 10 mV at 2 K for one of our measured devices. The hysteresis loops were observed in both the voltage-conductance curve and voltage-capacitance curve, which happens exactly in the same electrical field regime. Thus we can conclude that the capacitance change is directly related to the presence of the moving CDW under external electrical field. As shown in Fig. 4, the only difference is that the conductance is enhanced with the appearance of the moving CDW under the electrical field. However, the capacitance decreases with the existence of the moving CDW. Considering the two point contacting device as an electric capacity, it will not be a good electric capacity since it has some leakages. There will be a larger leakage with the presence of the moving CDW since a large current can pass through the device. The relation between the capacity and the applied voltage can be written down as $C=Q/V$, where C is the capacity of the device and Q is the stored electron charges with applied the external electrical field. Then the number of stored charges at 2 K just before and after the forward transition voltage are $2.56\times 10^{-10}$ C and $1.52\times 10^{-10}$ C, respectively. Thus the voltage-capacitance at low ac excitations can also been used to define the threshold voltage of CDWs, which give us a new route to investigate the few layered CDW devices.

**Conclusion:**

In summary, we carefully investigated the transport and capacitance properties of the ultrathin 2H-TaS$_2$ devices, where the nonlinear rather than linear current voltage characteristic was observed at low temperatures. The transition temperature of the charge density wave varies from device to device, which is attributed to the inhomogeneous local pinning centers of these micro-devices based on ultrathin 2H-TaS$_2$ flakes. However, the normalized temperature dependence of the threshold voltages depinned the CDW follows the same scaling law, which is proportional to $(1-T/Tr)^{0.5+\delta}$ with $\delta = 0.08$. The voltage-conductance and the voltage-capacitance at low ac excitations were measured simultaneously. The threshold voltage obtained from the voltage conductance is the same as that from the voltage-capacity measurements. We found that the voltage-capacity measurements can also be used to define the threshold voltage of the CDWs, which give us a new tool to investigate the CDWs.

This work was supported by "973 Program" Nos. 2011CB922200, 2014CB643903, 2011CBA00111 NSFC Grant Nos. 11174272, and 61225021. K.Y.W. acknowledges the support of Chinese Academy of Sciences "100 talent program". W.J.L acknowledges the support of the Joint Funds of the National Natural Science Foundation of China and the Chinese Academy of Sciences' Large-scale Scientific Facility (Grand No. U1232139). L.J.L acknowledges the support of the Director's Fund under Contract No.YZJJ201311 of Hefei Institutes of Physical Science, Chinese Academy of Sciences.

**Figure Captions:**

**Fig.1.** (a) AFM image of the typical exfoliated thinnest few layer TaS$_2$ with thickness of 2.11 nm. (b) the schematic diagram of the device structure; (c) The optical image of one of our fabricated devices.

**Fig.2.** (a) The current-voltage loop characteristics for both device #A(red) and #B(black) in at 2 K.Inset shows the I-V characteristics of device #A at different temperatures. (b) I-V loop characteristic for device #A in the voltage range between 0.48 to 0.55 V at temperatures $T = 2$ K (black), 20 K (cyan),25 K (magenta) ,30 K (dark yellow),35 K (purple). Inset shows the linear I-V characteristic of room temperature. (c) I-V loop characteristic for device #B in the voltage range between 0.48 to 0.55 V at temperatures $T = 2$ K (black), 20 K (cyan), 40 K (magenta ) , 50 K (dark yellow),60 K (purple).

**Fig.3.** (a) Temperature dependence of the $\Delta V$ for device #A (open circle) and #B (solid square). (b) The Normalized temperature dependence of the $\Delta V/(V_{2K}-V_{Tr})$ for device #A (open circle) and #B (solid square), where the red line is the fitting curve.

**Fig.4.** The voltage-conductance (black solid circle) and voltage-capacitance (red open circle) with 5 MHz ac excitation $V_{ac} = 10$ mV at 2 K.

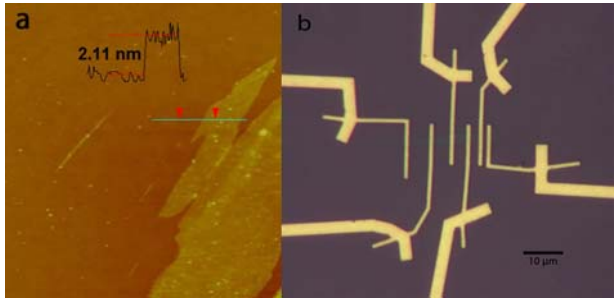

**Fig.1. Y. F. Cao et al.**

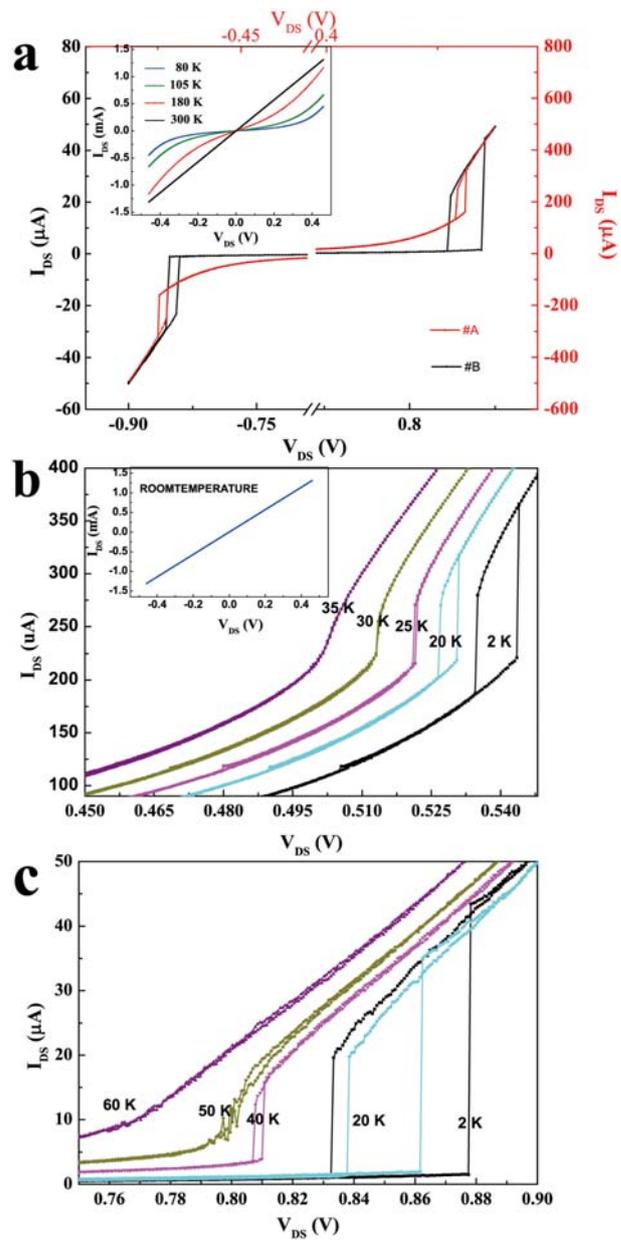

**Fig.2. Y. F. Cao et al.**

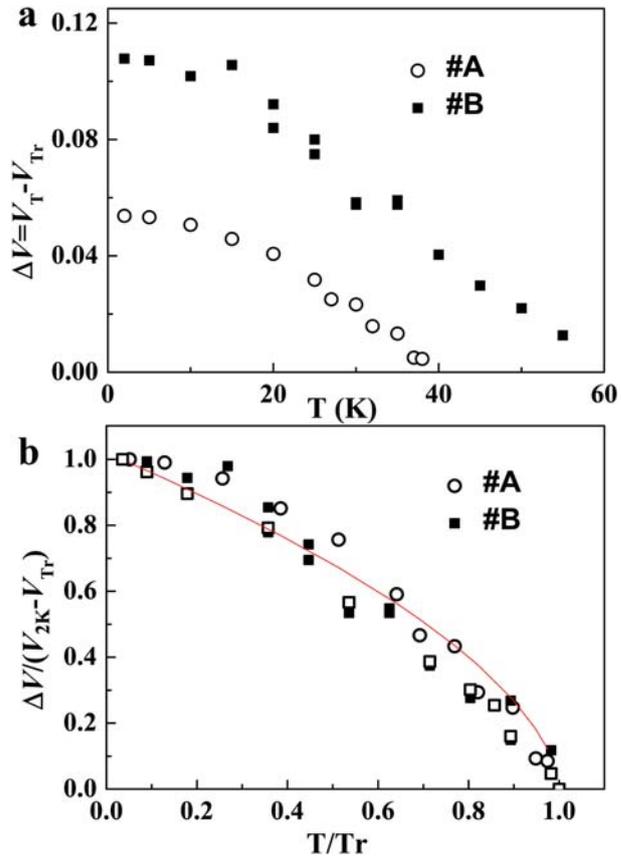

**Fig.3. Y. F. Cao et al.**

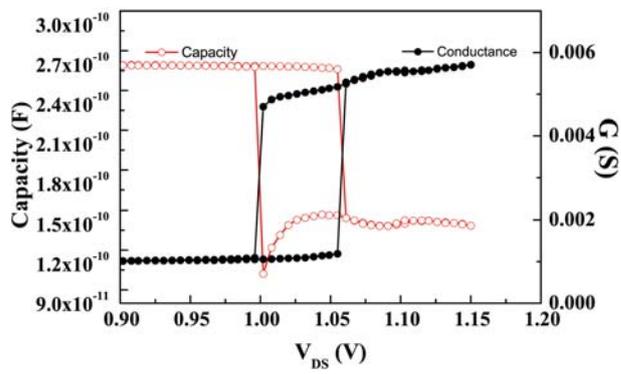

**Fig.4. Y. F. Cao et al.**